\documentclass[prd,aps,tightenlines,preprintnumbers,amsmath,amssymb,showpacs,superscriptaddress,nofootinbib,twocolumn,cites]{revtex4-1}
\usepackage{graphicx}
\usepackage{dcolumn}
\usepackage{bm}
\usepackage{hyperref}
\usepackage{placeins}

\def \beq{\begin{equation}}
\def \eeq{\end{equation}}
\def \bea{\begin{align}}
\def \eea{\end{align}}

\def\lsim{\mathrel{\rlap{\lower4pt\hbox{\hskip1pt$\sim$}}
    \raise1pt\hbox{$<$}}}                % less than or approx. symbol
\def\gsim{\mathrel{\rlap{\lower4pt\hbox{\hskip1pt$\sim$}}
    \raise1pt\hbox{$>$}}}                % greater than or approx. symbol

\newcommand{\tev}{\,\, \mathrm{TeV}}
\newcommand{\gev}{\,\, \mathrm{GeV}}

\newcommand{\ifb}{\ensuremath{\,\mbox{fb}^{-1}}}

\newcommand{\tb}{\tan\beta}
\newcommand{\MA}{m_A}
\newcommand{\Mh}{m_h}
\newcommand{\cp}{{\cal CP}}

\begin{document}

\preprint{ANL-HEP-PR-10-63}
\preprint{DESY 10--155}
\preprint{EFI-10-28}
\preprint{FERMILAB-PUB-10-474-T}

\title{Probing the Higgs Sector of High-Scale SUSY-Breaking Models at the
  Tevatron} 

\author{Marcela~Carena$^{1,2}$, Patrick~Draper$^{2}$, Sven~Heinemeyer$^{3}$, 
  Tao~Liu$^{2,4}$, Carlos~E.M.~Wagner$^{2,5,6}$, Georg~Weiglein$^{7}$\\[0.2cm]
\small\emph{$^1$~Fermi National Accelerator Laboratory, P.~O.~Box 500,
  Batavia, IL 60510, USA}\\ 
\small\emph{$^2$~Enrico Fermi Inst., Univ. of Chicago, 5640 S. Ellis Ave.,
  Chicago, IL 60637, USA}\\ 
\small\emph{$^3$~Instituto de F\'{\i}sica de Cantabria (CSIC-UC), E--39005
  Santander, Spain}\\ 
\small\emph{$^4$~Department of Physics, Univ. of California, Santa Barbara, 
  CA 93106, USA}\\
\small\emph{$^5$~HEP Division, Argonne National Laboratory, 9700 Cass
  Ave., Argonne, IL 60439, USA}\\ 
\small\emph{$^6$~KICP and Dept. of Physics, Univ. of Chicago, 5640 S. Ellis
  Ave.,Chicago IL 60637, USA}\\ 
\small\emph{$^7$~DESY, Notkestrasse 85, D--22607 Hamburg, Germany}\\
}

\date{\today}

\begin{abstract}
A canonical signature of the Minimal Supersymmetric Standard Model
(MSSM) is the presence of a neutral Higgs boson with mass bounded from
above by about $135 \gev$ and Standard Model (SM)-like couplings to the 
electroweak gauge bosons.
In this note we investigate the reach of the Tevatron collider for the MSSM
Higgs sector 
parameter space associated with a variety of high-scale minimal models
of supersymmetry (SUSY)-breaking, including the Constrained MSSM
(CMSSM), minimal Gauge Mediated SUSY-breaking (mGMSB), and minimal
Anomaly Mediated SUSY-breaking (mAMSB).
% In each case the lightest Higgs boson possesses Standard Model
% (SM)-like couplings to the electroweak gauge bosons.  
We find that the Tevatron can provide strong constraints on these models via Higgs boson searches.
%With an integrated luminosity of $16 \ifb$
%per detector and an efficiency improvement of $\sim 20\%$ compared to the present situation, we project that 
%%\marginpar{{\small GW: changed $\sim 30\%$ to $\sim 20\%$ here, since 
%%$10\%$ have already been achieved; same for $50\% \to 40\%$}}
%these models could completely be ruled out by the Higgs searches at the
%Tevatron. With 40\% analysis improvements and $16 \ifb$, evidence 
%at the $3\sigma$ level for the light Higgs boson could be
%expected in some regions of parameter space.
Considering a simple projection for the efficiency improvements in the Tevatron analyses, we find that with an integrated luminosity of $16\ifb$ per detector and an efficiency improvement of  20\% compared to the present situation, these models could be probed essentially over their entire ranges of validity.
%, by the Higgs searches at the Tevatron. 
With 40\%
analysis improvements and $16 \ifb$, our projection shows that evidence at the $3\sigma$ level for the light Higgs boson could be expected in extended regions of parameter space.

% Our results should be considered conservative in the sense that the total set of analysis improvements planned by the collaborations amounts to 20\% more than the maximum we apply in the low mass region.
\end{abstract}
%\pacs{12.60.Fr, 13.20.Gd, 13.66.Hk, 14.80.Cp}

\maketitle
\renewcommand{\thepage}{\arabic{page}}

%%%%%%%%%%%%%%%%%%%%%%%%%%%%%%%%%%%%%%%%%%%%%%%%%%%%%%%%%%%%%%%%%%%%%%%%%%%%%%
%%%%%%%%%%%%%%%%%%%%%%%%%%%%%%%%%%%%%%%%%%%%%%%%%%%%%%%%%%%%%%%%%%%%%%%%%%%%%%

\section{Introduction}
The Tevatron collider at Fermilab is now in its twenty-sixth year of
$p\bar{p}$ collisions. Operating at a center-of-mass energy of 1.96~TeV,
it is a highly productive and well-understood machine, and the 
rate of the 
%{\bf GW: o.k.? PD: changed ``accelerate" to ``increase"}
delivered
luminosity continues to increase.  Nonetheless, with the successful
collision of protons at a center-of-mass energy of 7~TeV at the Large
Hadron Collider (LHC), the position at the energy frontier held by the
Tevatron thus far is now being taken over by the LHC. According to the current
schedule, it is planned to run the Tevatron until the end of 2011, 
while recently the Fermilab Physics Advisory Committee has given the
recommendation to extend the operation of the Tevatron until the end of 
2014~\cite{TPAC}. It is therefore of interest to assess the physics reach
achievable with the final Tevatron dataset, based on the scenarios of
running until the end of 2011 or 2014 (for a recent summary, see~\cite{Wood:2010vs}.)

One of the most important ongoing tasks of the Tevatron is the search for new 
particles directly associated with electroweak symmetry breaking (EWSB).
Searches for the Standard Model (SM) Higgs boson by the CDF and
D\O~collaborations have excluded the fundamental scalar in
the mass range $158$--$175 \gev$ at 95\% C.L. via the $W^+W^-$ decay
channel~\cite{:2010ar}. The sensitivity in that region can still continue
to grow as the volume of analyzed data increases. However, it is in the
mass range below $135 \gev$ where the Tevatron achieves perhaps its greatest
relevance.  In this regime, Higgs production in association with an
electroweak gauge boson provides the most sensitive search
channels. Associated production clearly demonstrates the 
relationship of the scalar to the mechanism of electroweak symmetry
breaking. Furthermore, access to the coupling of a light Higgs boson to
bottom quarks will be a crucial input in determining the Higgs couplings
to fermions and gauge bosons~\cite{Duhrssen:2004cv,Lafaye:2009vr} and will thus be
essential for establishing the Higgs mechanism as a whole. The
experimental information achievable at the Tevatron will be
complementary to the results of the Higgs searches at the LHC, where the
low-mass region below $\sim 135 \gev$ turns out to be the most
challenging one~\cite{Aad:2009wy,Ball:2007zza}. At the LHC the
accumulation of a significant dataset, of the order of 
$10 \ifb$~\cite{lhc2fc}, would be necessary for the discovery of a
SM-type Higgs boson in the $gg \to h \to \gamma\gamma$ channel. Further
information at the LHC can be expected from weak boson fusion Higgs
production and eventually, with sufficient luminosity, also from the
associated production channels. There exists also the exciting
possibility that important Higgs production channels at the 
LHC could arise from
decays of states of new physics, such as supersymmetric particles.
The LHC and the Tevatron will complement each other in the search for a
light SM-like Higgs by exploiting different channels and different
background levels, especially in the $b \bar b$ final states where the 
different nature of $pp$ vs.\ $p \bar p$ collisions becomes relevant.

Based on the current data acquisition rate, running the Tevatron until
the end of 2011 or 2014
will yield approximately $10$ or $16 \ifb$ of analyzable data per
experiment, respectively.  Furthermore, a number of analysis improvements
for the Higgs searches are ongoing, and the collaborations estimate that
on the order of 50\% improvements (with respect to the status of March
2009) are achievable~\cite{fishertalk,CDFSM}.

Low-scale minimal supersymmetry (SUSY) is a well-motivated theory of new
electroweak-scale physics that resolves a number of open problems in the
SM. In addition to providing a technical solution to the hierarchy
problem, it offers a viable weakly-interacting dark matter candidate,
exhibits gauge coupling unification at high scales, and generates EWSB
via radiative effects~\cite{reviews}. 

If it exists in nature, SUSY must be broken.  Flavor experiments
strongly suggest that the breaking of SUSY must be communicated to the
MSSM fields in an approximately flavor-diagonal
%\marginpar{{\small GW: don't like phrase ``flavor-blind''}}
manner~\cite{Ellis:1981ts,Bertolini:1990if,Isidori:2001fv,Buras:2002vd,Babu:1999hn,Dedes:2002er}. 
Three ``standard'' high-scale models with flavor-universal SUSY-breaking
parameters are the Constrained MSSM (CMSSM)~\cite{Hall,mSUGRArev},
minimal Gauge Mediated SUSY-breaking
(mGMSB)~\cite{Giudice:1998bp,Affleck:1984xz,Dine:1995ag} and minimal
Anomaly Mediated SUSY-breaking (mAMSB)~\cite{Randall:1998uk,Giudice:1998xp}. 
%Within the CMSSM all the soft SUSY-breaking scalar
%masses are assumed to take a universal value $m_0$ at the GUT scale, as are the
%soft SUSY-breaking gaugino masses $m_{1/2}$ and trilinear couplings
%$A_0$. 
Within the
CMSSM  the soft SUSY-breaking scalar masses are
assumed to take a universal value $m_0$ at the GUT scale, while the 
 soft SUSY-breaking gaugino masses take a GUT universal value of $m_{1/2}$ and
the trilinear couplings take a common GUT value $A_0$, respectively.
In mGMSB, SUSY is broken in a hidden sector that affects the MSSM only
through gauge interactions, mediated via ``messenger'' particles, 
leading automatically to flavor-universal soft parameters in the
effective theory. 
Here $M_{\rm mess}$ denotes the overall messenger mass scale; 
$N_{\rm mess}$ is a number called the 
messenger index, parameterizing the structure of the messenger
sector, and $\Lambda$ is the universal soft SUSY-breaking mass scale
felt by the low-energy sector.
Finally, in mAMSB the soft terms are generated by
the superconformal anomaly. The overall scale of SUSY particle masses is
set by $M_{\rm aux}$, which is the vacuum expectation value 
of the auxiliary field in the supergravity multiplet.
Furthermore,  a phenomenological parameter $m_0$ is introduced
to avoid negative squared slepton mass squares.  In all three models,
the high-scale parameters are supplemented by $\tb$ and sign$(\mu)$ as
additional inputs (see below).

At low scales, the MSSM Higgs sector exhibits rich
phenomenology. There are five physical states: the light and heavy
$\cp$-even Higgs bosons\footnote{We concentrate here on the case without
  $\cp$-violation.}, $h$ and $H$, the $\cp$-odd Higgs $A$, as well as
the two charged states, $H^\pm$~\cite{hhg}.
The lightest Higgs boson behaves SM-like for $\MA \gsim 150 \gev$, and
its mass has an upper limit of 
$\Mh \lsim 135 \gev$~\cite{Degrassi:2002fi}. 
In the limit of large $\MA$ the $H$ has negligible couplings of the form
$VVH$ to gauge bosons, whereas the $A$ has 
vanishing $VVA$ couplings (at tree-level). Both heavy neutral Higgs
bosons have $\tb$-enhanced couplings to down-type fermions.
The tree level couplings and
masses of the Higgs bosons are controlled entirely by two parameters,
which can be taken to 
be the $\cp$-odd mass $\MA$, and $\tb$, the ratio of the vacuum
expectation values of the neutral components of the two Higgs
doublets. Radiative corrections introduce dependence on other MSSM
parameters. This dependence is dominated by the stop masses and the stop mixing
parameter $X_t = A_t-\mu\cot\beta$, where $\mu$ denotes the Higgs mixing
parameter and $A_t$ is the stop soft trilinear coupling. 

The high-scale models typically generate a SM-like lightest
$\cp$-even Higgs state~\cite{Ellis:2001qv,Ambrosanio:2001xb,Heinemeyer:2008fb}. 
Consequently, LEP and Tevatron searches for the SM Higgs boson can be
applied to the case of the lightest Higgs $\cp$-even Higgs boson in
those supersymmetric models. 
Conversely, in these models $H$ and $A$ tend to have negligible couplings to
SM gauge bosons, and different searches must be used.

%The Tevatron reach for the SM Higgs can be
%rescaled channel-by-channel to obtain the reach for each Higgs state in
%the MSSM at each point in parameter space. An estimated maximal
%statistical significance for each MSSM point is thus achieved by
%combining all channels for all states in quadrature. 
%LATER

In this work we will examine the projected capabilities of the Tevatron
to provide evidence for or exclude regions of the MSSM ($\MA$,$\tb$,$\Mh$)
%($\MA$,$\tb$) plane 
parameter space of these three soft SUSY-breaking scenarios. 
We begin in Section II
with a brief review of the calculation of combined expected statistical
significances from Higgs searches. For comparison with the MSSM reach,
we present the expected significance 
for SM Higgs searches in our approach in Section III.
%and discuss the extension to the MSSM. 
In Sections
IV-VI we give the projected reach for each high-scale SUSY-breaking model. 
In Section VII we conclude.

%%%%%%%%%%%%%%%%%%%%%%%%%%%%%%%%%%%%%%%%%%%%%%%%%%%%%%%%%%%%%%%%%%%%%%%%%%%%%%
%%%%%%%%%%%%%%%%%%%%%%%%%%%%%%%%%%%%%%%%%%%%%%%%%%%%%%%%%%%%%%%%%%%%%%%%%%%%%%

\section{Expected Combined Significances}

%To make our projections in general we follow exactly the approach as in 
To make our projections we follow the approach used in 
Refs.~\cite{Draper:2009au,Draper:2009fh}. 
We take as input the March 2009 expected limits
%\marginpar{{\small explain / justify?}}
on Higgs signals from a number of different search channels given by the
CDF and $\mbox{D\O}$ experiments in
Refs.~\cite{CDFSM,D0SM,D0comb,tautau,Collaboration:2009zh,cdfcharged}. 
The main purpose in considering the 2009 expected limits rather than the 2010 limits
is to keep consistent the meaning of efficiency improvements with respect to 
Refs.~\cite{Draper:2009au,Draper:2009fh}, as we will discuss further below. 
Since the expected limits change primarily through efficiency and luminosity
increases, the results should be insensitive to whether 2009 or 2010 limits 
are used as a baseline, so long as the efficiency improvements are defined relative to
that baseline.
It is important for projections to consider the expected reach rather than
the observed, as the latter may contain fluctuations that should not be
assumed in future data.

We take into account the dominant search channels for each type of Higgs
boson. 
Tevatron searches for the SM-like neutral Higgs are performed mainly in
two channels: gluon fusion production with Higgs decay to $W^+W^-$,
which is most sensitive to Higgs masses above $135 \gev$, and in the
associated production with a $W$ or $Z$ boson and subsequent decay to
$b\bar{b}$, which contributes most prominently below
$135 \gev$~\cite{Abazov:2008eb,Aaltonen:2008zx,Abazov:2005un,Aaltonen:2008ec}. 
The neutral Higgs states with small gauge couplings (``nonstandard Higgs
bosons'')  are efficiently produced primarily for relatively low $\MA$ and large
$\tb$ through gluon fusion via a loop of bottom quarks, or in associated
production with bottom quarks, and then decay dominantly to $b\bar{b}$
and $\tau^+\tau^-$.  The inclusive search for $\tau^+\tau^-$ final
states and the exclusive searches for $b\tau^+\tau^-$ and $3b$ final
states provide the dominant sensitivity for nonstandard Higgs bosons at the
Tevatron~\cite{Balazs:1998nt,Baer:1998rg,Drees:1997sh,Carena:1998gk,Belyaev:2002zz,:2008hu,Abulencia:2005kq,Abazov:2008zz,CDF3b,btautau,tautaucomb}. Finally,
the main search channel for a charged Higgs boson (if it is lighter than
the top quark) at the Tevatron is 
$t\to H^+b,\; H^+\to\tau^+\nu$ for $\tb>1$, and reaches maximal effect
in the same parameter region as the nonstandard Higgs
searches~\cite{Collaboration:2009zh,Abulencia:2005jd}.

\medskip
We begin our MSSM analysis with scans over the high-scale input parameters of
the CMSSM, mGMSB, and mAMSB models. The resulting MSSM soft
parameters are run down to the electroweak scale with the code
SoftSUSY~\cite{softsusy}, and experimental bounds on the lightest
neutralino and chargino masses are applied~\cite{Amsler:2008zzb}. For
points that are consistent with these bounds, the parameters are fed
into the code
FeynHiggs~\cite{Heinemeyer:1998yj,Heinemeyer:1998np,Degrassi:2002fi,Frank:2006yh} 
to compute the Higgs spectrum at the 2-loop order, as well as the couplings and branching ratios. Following the procedure in
Refs.~\cite{Draper:2009au,Draper:2009fh}, the Higgs sector observables
are then used to convert the expected limits on the SM Higgs given by the
%\marginpar{{\small added: ``expected''}}
experimental collaborations as well as expected limits on non-SM-like Higgs
bosons into signal significances in the CMSSM, mGMSB, and mAMSB
parameter spaces. As input value for the top quark mass we use in our
%\marginpar{{\small new text; add remark on treatment of theoretical uncertainties?}}
analysis $m_t = 173.1 \gev$.  The latest experimental value for $m_t$ is~\cite{mt1733}
\begin{align}
m_t^{\rm exp} &= 173.3 \pm 1.1 \gev~.
\label{mtexp}
\end{align}
Taking into account the experimental uncertainty at the $2\sigma$ level
and adding the difference of $0.2 \gev$ between the current experimental
value and the one used in our analysis, the calculated value of $m_h$
could move upwards by $\sim 1.5 \gev$ (see Table 4.1 in Ref.~\cite{PomssmRep}
for details.) Furthermore, the theoretical uncertainty from unknown
higher-order corrections in the prediction for the light $\cp$-even
Higgs mass in the CMSSM, mGMSB and mAMSB scenarios can be estimated to
be about 1--$2\gev$~\cite{Heinemeyer:2008fb}. 
Combining the parametric uncertainty from $m_t$ quadratically with a theory uncertainty of $\sim 1.5 \gev$
results in a possible upward shift of $m_h$ of up to $2.2 \gev$.
We discuss below the effects of these uncertainties.

In the $B \gg S$ ($B$=background, $S$=signal) Gaussian approximation, the significances scale with
the square root of the luminosity and linearly with increases in signal
efficiency. This approximation was shown in Ref.~\cite{Draper:2009au}
 to match well with more precise combinations of Tevatron Higgs limits, and we
use it in the present work. We analyze the 
Tevatron potential for the cases of $10 \ifb$ and $16 \ifb$ of final
integrated luminosity per experiment, and 10\% to 50\% efficiency
improvements. 
We emphasize that the analysis improvements examined in this work are
taken with respect to the March 2009 Tevatron expected limits
(following exactly the approach in
Refs.~\cite{Draper:2009au,Draper:2009fh}). A comparison of the projections for the SM Higgs given in
Ref.~\cite{Draper:2009au} with the expected limits for
$\approx 6$ fb$^{-1}$ presented in Ref.~\cite{:2010ar} indicates that
effectively 10\% improvements have already been achieved in the low-mass
region and 20\% improvements have been achieved in the high-mass
region between March 2009 and Summer 2010. As the low-mass searches are 
most relevant for the SM-like Higgs in the MSSM, the universal 10\%
improvement gives an estimated MSSM reach assuming \textit{no further}
improvements beyond what has been already achieved by the summer 2010.

In our figures we present the estimated maximal reach for
any given point in model space, obtained by combining in quadrature the
expected significance for each individual search channel, including
those that search for SM-like, nonstandard, and charged Higgs bosons
from both CDF and D\O.  In the Gaussian
approximation, $S/\sqrt{B}$ can be interpreted either as the exclusion
or the discovery power.
In other words, a point marked as ``n$\sigma$'' can be excluded at 
%\marginpar{{\small rephrased}}
the n$\sigma$ level if the signatures in the Higgs sector corresponding 
to this point are excluded at this level of confidence, or one
could obtain an n$\sigma$ ``evidence'' if signatures compatible with a
corresponding signal were observed.
For reference we also shade LEP-excluded
regions as derived from Ref.~\cite{Schael:2006cr}.
For each soft SUSY-breaking scenario we also present the minimum required
improvement in efficiency for all points to be probed 
at the $2\,\sigma$ level (corresponding to an exclusion at the
95\%~C.L.), once a total integrated luminosity of $16 \ifb$ per 
experiment is analyzed.

\begin{figure*}
\begin{minipage}{1.0\linewidth}
\begin{center}
\includegraphics[width=0.8\textwidth]{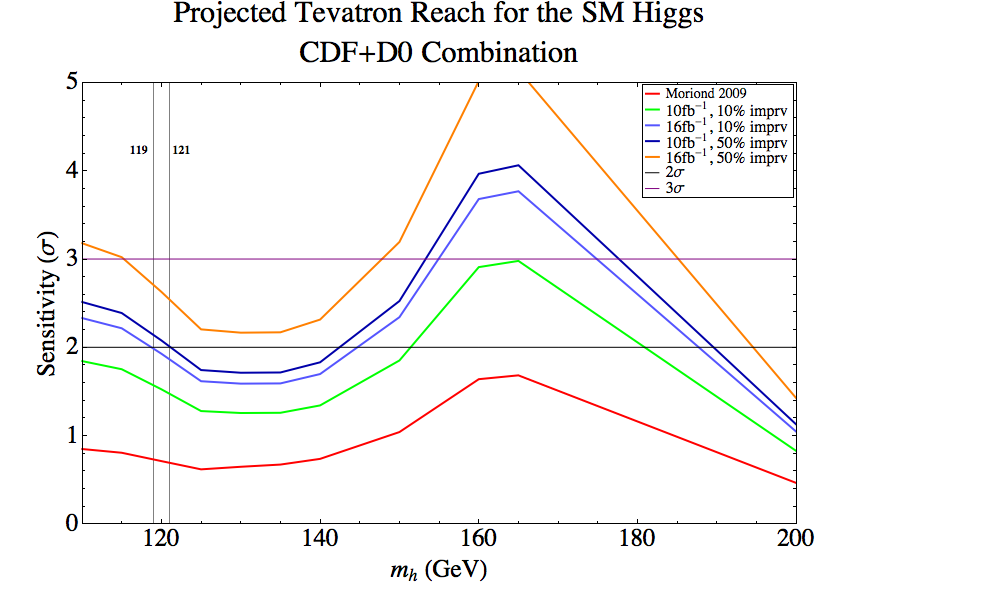}
\caption{Projected Tevatron coverage of SM Higgs masses
for a range of final luminosities and efficiency improvements. The curve labelled ``Moriond 2009''
   indicates the March 2009 expected limits.}
\label{SMfig}
\end{center}
\end{minipage}
\end{figure*}

%%%%%%%%%%%%%%%%%%%%%%%%%%%%%%%%%%%%%%%%%%%%%%%%%%%%%%%%%%%%%%%%%%%%%%%%%%%%%%

\section{The SM}

Before analyzing the Tevatron reach for the MSSM Higgs sector, it is useful to briefly review the reach for the SM Higgs, which can then be used to understand some features of the MSSM plots in the subsequent sections. In Fig.~\ref{SMfig} we give the projections for the two luminosity and efficiency
 improvement assumptions as a function of $m_h$.

Three features are immediately apparent. Firstly, for low $m_h\sim 115$ GeV, the Tevatron is expected to achieve 3$\sigma$ sensitivity with $16 \ifb$ and $50\%$ analysis improvements. Secondly, even with 50\% improvements, more than $10 \ifb$ is necessary to cover the entire low mass range at more than the $2\sigma$ level. Finally, for the high mass range, 3$\sigma$ sensitivity is expected in a broad range, from $185$ GeV to below $150$ GeV, with $16 \ifb$ and $50\%$ analysis improvements.

We note that these projections are somewhat weaker than those presented in~\cite{run3} and have a slightly different shape in the low-mass region.  The primary reason for the difference is that we have assumed a rate of efficiency improvement
independent of the Higgs mass, while the Tevatron experiments expect additional gains in efficiency
improvement in certain mass regions. In particular, they expect efficiency
improvements larger than 50\% (40\% with respect to today's analyses) to be possible in the
mass region between 120 and 140 GeV (see, e.g.,~\cite{denisovPAC}).  In our analysis we use flat improvement factors, as
presented in Fig.~1, since they allow a simpler analytical treatment and understanding of the Higgs 
reach projections. Since in most of the parameter space, the MSSM Higgs reach is controlled by SM 
Higgs searches, the extrapolation of our results to different values of the efficiency improvements can 
be performed in a straightforward way using the curves presented in Fig.~1. We expect that a detailed mass-dependent implementation of the efficiency improvements for the low Higgs mass region, as presented in~\cite{denisovPAC}, will lead to an expected significance of $3\sigma$ with $16\ifb$ over nearly the whole parameter space in the models under study.

%The primary reason for the difference is that our 50\% efficiency improvement assumption (40\% with respect to today's analyses) is less than what may be achieved (see, e.g.,~\cite{denisovPAC}). We use our improvement factors as a more conservative estimate of the Tevatron's final median sensitivity.

%%%%%%%%%%%%%%%%%%%%%%%%%%%%%%%%%%%%%%%%%%%%%%%%%%%%%%%%%%%%%%%%%%%%%%%%%%%%%%

\begin{figure*}
\begin{minipage}{1.0\linewidth}
\begin{center}
\begin{tabular}{cc}
\includegraphics[width=0.50\textwidth]{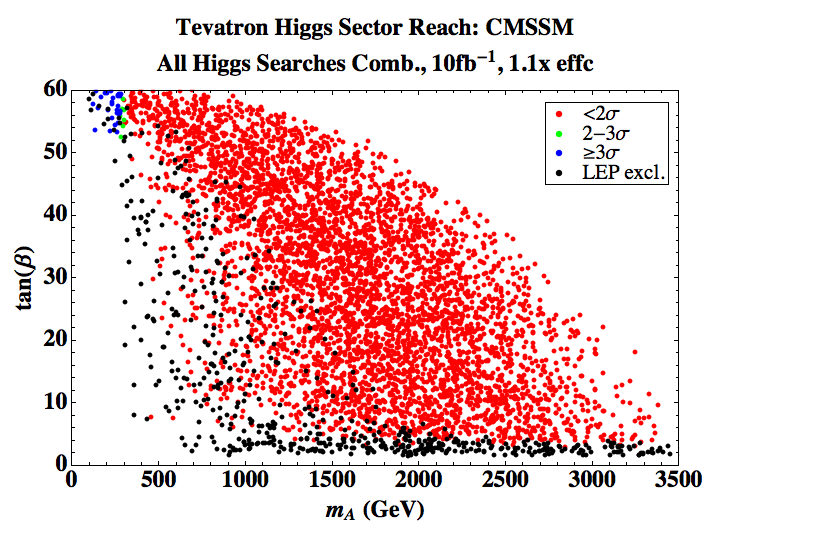} &
\includegraphics[width=0.50\textwidth]{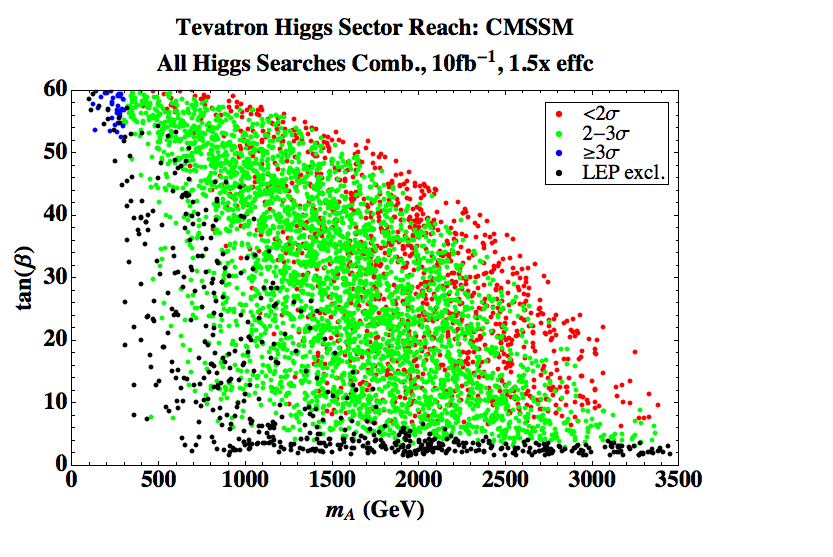} \\
\includegraphics[width=0.50\textwidth]{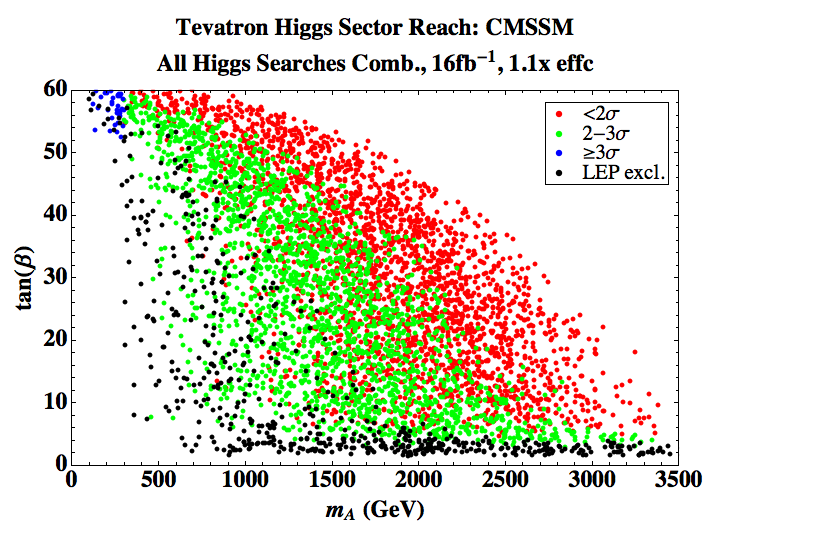} &
\includegraphics[width=0.50\textwidth]{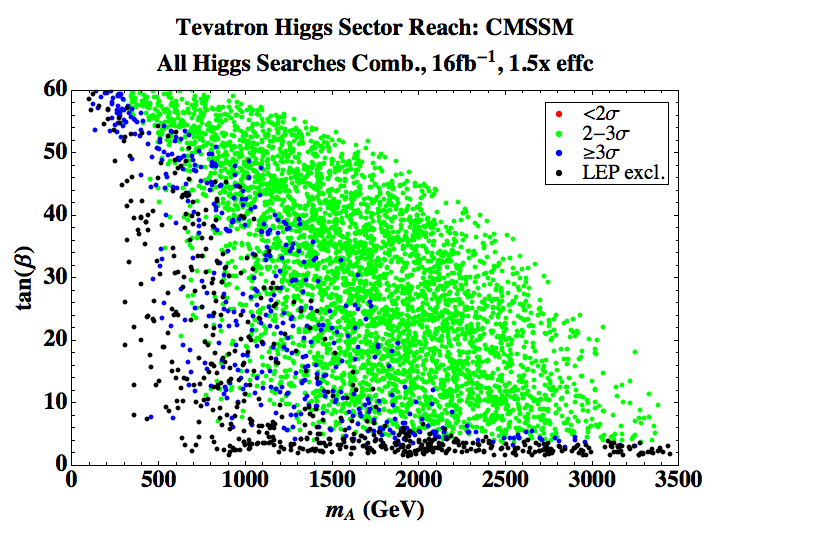}
\end{tabular}
\caption{Projected Tevatron coverage of the CMSSM on the ($\MA$, $\tb$)
  plane with $10 \ifb$ and 10\% efficiency improvement (\textit{top
    left}),  $10 \ifb$ and 50\% improvement (\textit{top right}),
  $16 \ifb$ and 10\% efficiency improvement (\textit{bottom left}),
  and $16 \ifb$ with 50\% improvement (\textit{bottom right}). The efficiency improvements are given
 relative to the March 2009 expected limits.} 
\label{CMSSMfig}
\end{center}
\end{minipage}
\end{figure*}

\begin{figure*}
\begin{minipage}{1.0\linewidth}
\begin{center}
\includegraphics[width=0.5\textwidth]{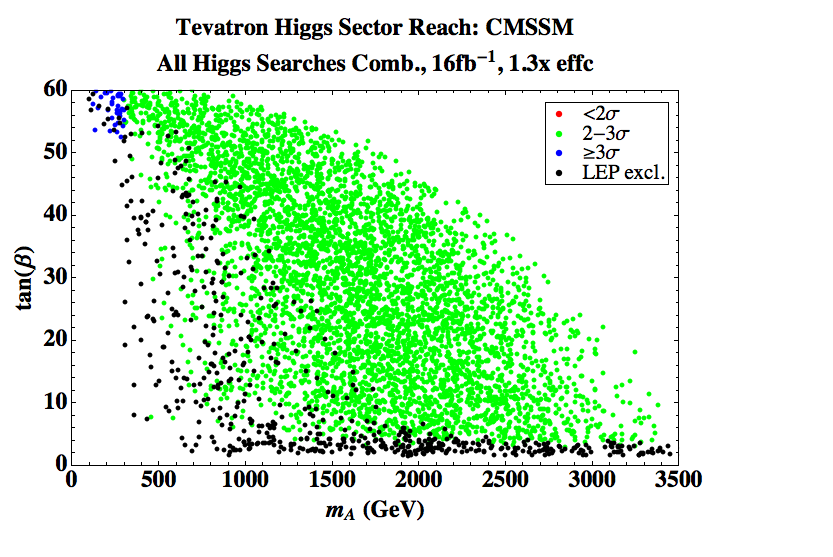}
\caption{Projected Tevatron coverage of the CMSSM on the ($\MA$, $\tb$)
  plane with $16 \ifb$ and 30\% efficiency improvement (w.r.t. March
 2009), which is the
  approximate threshold for full coverage at the 2$\sigma$ level.}  
\label{CMSSMfig2sig}
\end{center}
\end{minipage}
\end{figure*}

\begin{figure*}
\begin{minipage}{1.0\linewidth}
\begin{center}
\begin{tabular}{cc}
\includegraphics[width=0.50\textwidth]{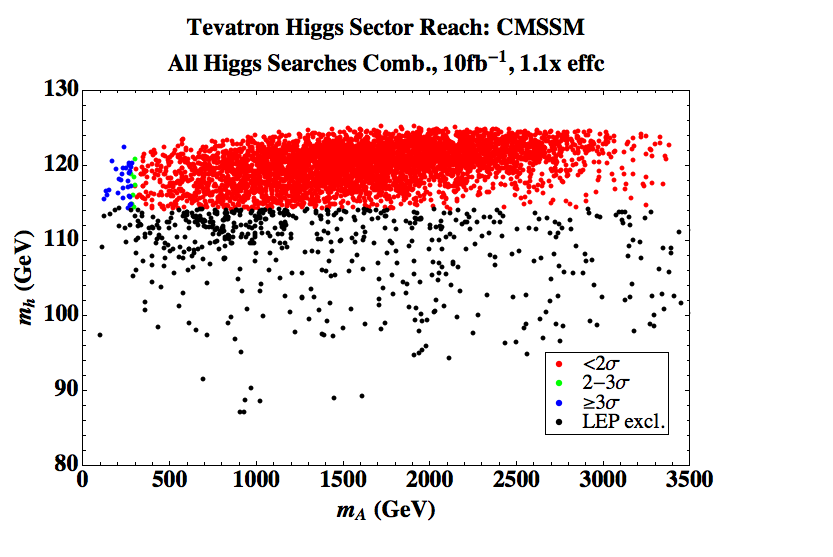} &
\includegraphics[width=0.50\textwidth]{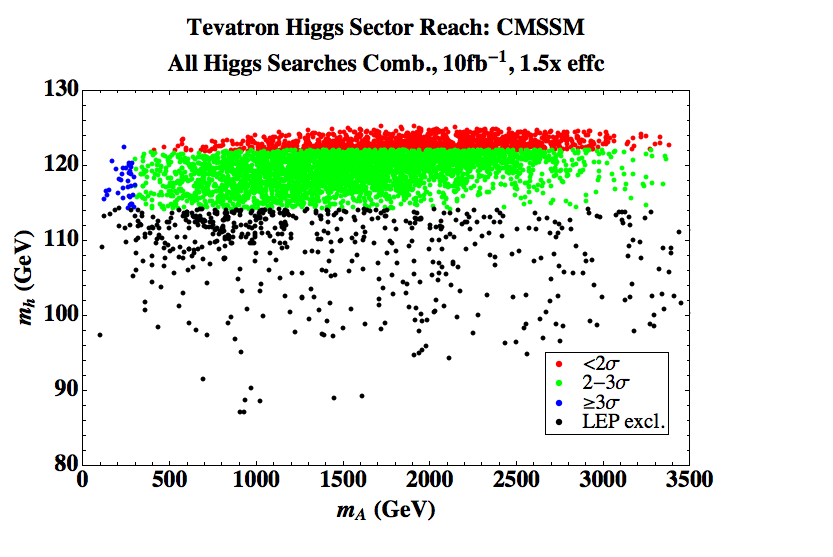} \\
\includegraphics[width=0.50\textwidth]{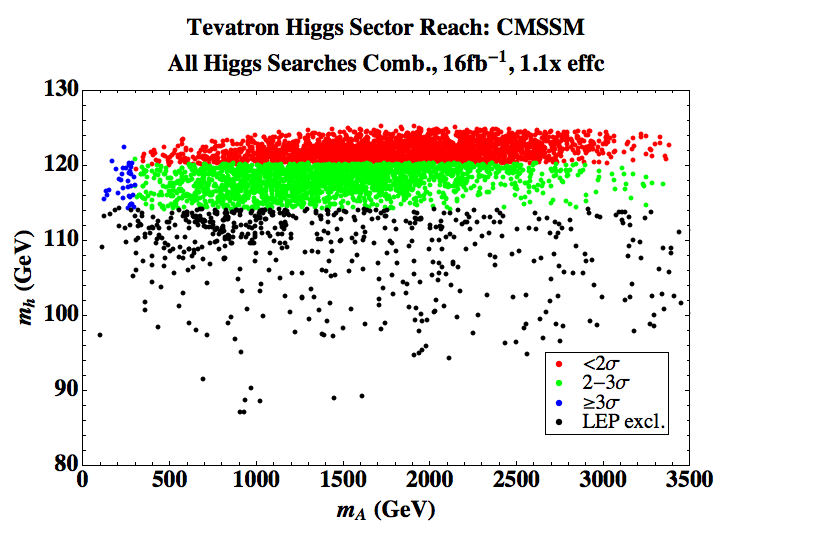} &
\includegraphics[width=0.50\textwidth]{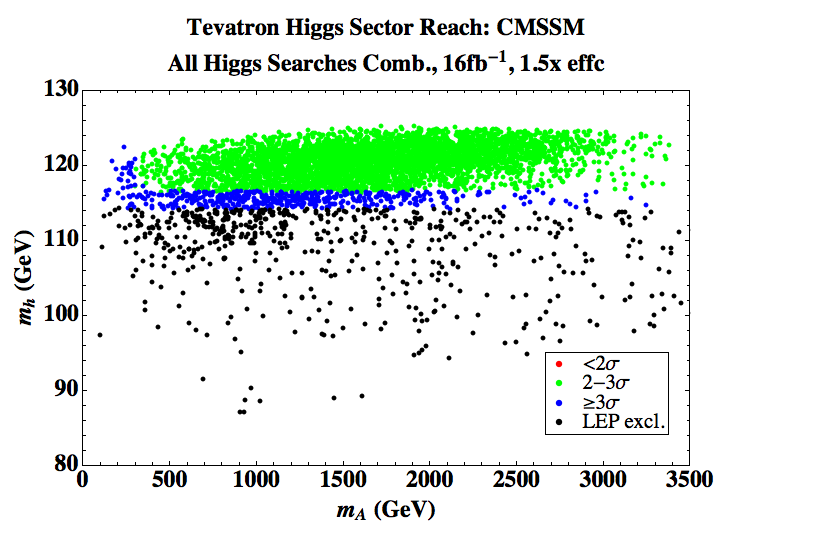}
\end{tabular}
\caption{Projected Tevatron coverage of the CMSSM on the ($\MA$, $\Mh$)
  plane with $10 \ifb$ and 10\% efficiency improvement (\textit{top
    left}),  $10 \ifb$ and 50\% improvement (\textit{top right}),
  $16 \ifb$ and 10\% efficiency improvement (\textit{bottom left}),
  and $16 \ifb$ with 50\% improvement (\textit{bottom right}). The efficiency improvements are given
 relative to the March 2009 expected limits.} 
\label{CMSSMfigmhma}
\end{center}
\end{minipage}
\end{figure*}

\section{The Constrained MSSM}

In the CMSSM, the messenger scale 
%of SUSY-breaking 
is taken to be the unification
scale, $\approx 2\times10^{16} \gev$. At this scale the sfermions and
Higgs bosons are assigned a common soft mass $m_0$, the gauginos share a
soft mass $m_{1/2}$, and the soft trilinear couplings are set to a value
$A_0$. The low-energy inputs are $\tb$ and the sign of $\mu$. We
fix the sign of $\mu$ to be positive to give a positive contribution to
the anomalous magnetic moment of the muon, $(g-2)_\mu$, thus not
worsening the SM
prediction~\cite{Moroi:1995yh,Bennett:2006fi,doink,Heinemeyer:2008fb}. 
The remaining parameters are scanned over the ranges
\begin{align}
50 \gev \leq m_0 \leq 2 \tev, &\quad 50 \gev\leq m_{1/2}\leq 2 \tev,\nonumber\\ 
-3 \tev \leq A_0\leq 3 \tev, &\quad 1.5 \leq \tb \leq 60.
\end{align}

%\begin{figure*}[t]
%\begin{minipage}{1.0\linewidth}
%\hspace{-1.5cm}
%\begin{center}
%\begin{tabular}{cc}
%\includegraphics[width=0.50\textwidth]{CMSSM_mhtb_lin_10ifb_1pt1x} &
%\includegraphics[width=0.50\textwidth]{CMSSM_mhtb_lin_10ifb_1pt5x} \\
%\includegraphics[width=0.50\textwidth]{CMSSM_mhtb_lin_16ifb_1pt1x} &
%\includegraphics[width=0.50\textwidth]{CMSSM_mhtb_lin_16ifb_1pt5x}
%\end{tabular}
%\caption{Projected Tevatron coverage of the CMSSM on the ($\tb$, $\Mh$)
%  plane with $10 \ifb$ and 10\% efficiency improvement (\textit{top
%    left}),  $10 \ifb$ and 50\% improvement (\textit{top right}),
%  $16 \ifb$ and 10\% efficiency improvement (\textit{bottom left}),
%  and $16 \ifb$ with 50\% improvement (\textit{bottom right}).} 
%\label{CMSSMfigmhtb}
%\end{center}
%\end{minipage}
%\end{figure*}

The resulting expected significances on the ($\MA$,~$\tb$) plane are 
given in Figs.~\ref{CMSSMfig} and~\ref{CMSSMfig2sig}, for values of the
luminosity and signal efficiency in each panel as given in the figure
captions. The results of Fig.~\ref{CMSSMfig} are also projected onto the
($\MA,\Mh$) plane in Fig.~\ref{CMSSMfigmhma}. The most important feature
can be seen in the lower right plot of Fig.~\ref{CMSSMfig}, where  
$16 \ifb$ and a 50\% increase in efficiency (compared to March 2009)
are assumed. If these
two improvements are achieved, a $3\sigma$ sensitivity to the SM-like Higgs 
is reached in parts of the parameter space. On the other hand, as shown
in Fig.~\ref{CMSSMfig2sig}, $16 \ifb$ and 30\% increase in sensitivity
is sufficient to achieve
a 2$\sigma$ sensitivity over the whole parameter space. Consequently,
the complete model could be excluded at the 95\% C.L.\ in this case, 
or it would yield
at least a $2\sigma$ ``excess'' in the Higgs boson searches.

Examining Figs.~\ref{CMSSMfig} and~\ref{CMSSMfig2sig} in more detail,
the scatter points exhibit a characteristic curve bounding the upper value of 
$\MA$ for a given value of $\tb$. This behavior can be understood from
the dependence of the low scale value of $\MA^2$ on the splitting
between two soft SUSY-breaking parameters in the MSSM Higgs sector,
$m_{H_u}^2$ and $m_{H_d}^2$, given by  
\begin{align}
\MA^2\approx m_{H_d}^2-m_{H_u}^2-m_Z^2
\end{align}
in the large $\tb$ limit~\cite{copw}. At the electroweak scale, the splitting
approaches zero as $\tb$ increases, because the $\tb$-enhanced bottom
Yukawa coupling approaches the top Yukawa coupling 
and uniformizes the RG evolution of
$m_{H_d}^2$ and $m_{H_u}^2$. The splitting generated by RG running is
also roughly proportional to the squark masses squared,
$m_{\tilde{Q}}^2$, times a logarithm, so the 
points that saturate the boundary curve are typically those for which
the squarks are heavy. Heavy squarks also generate a larger 
$\Mh$, so most of the CMSSM points near the boundary represent models
that can only be probed at the $2\sigma$ level with both $16 \ifb$ and
30\% improvements.  These are typically models with larger values of
$m_{1/2}$, which efficiently raises the low-scale squark masses, as well
as larger negative values of $A_0$, which further increases the SM-like
Higgs mass. Note that in such cases the squarks and gluino are typically beyond 
the kinematic reach of the LHC.
For low values of $\MA$, large $\tb$, and either $10 \ifb$ with 
50\% analysis improvements or $16 \ifb$ with only 
10\% analysis improvements, the nonstandard Higgs searches in the
$\tau\tau$ and $3b$ channels provide the only expected $3\sigma$
significance. In contrast, for $16 \ifb$ with 50\% improvements, models
with lighter squarks can give rise to a $3\sigma$ excess in the $b\bar{b}$
search channels for a light SM-like Higgs. 
%Furthermore, even for
%very large values of the squark masses, the Tevatron can exclude the
%\marginpar{{\small repetition! Remove last sentence of this par?}}
%CMSSM Higgs sector with the larger data set and 30\% improvement, as can
%be seen in Fig.~\ref{CMSSMfig2sig}.

To understand clearly the strong correlation between the expected reach
and $\Mh$,  in Fig.~\ref{CMSSMfigmhma} we plot the scan points on the
($\MA$,~$\Mh$) plane, with luminosity and efficiency improvements in
each panel as in Fig.~\ref{CMSSMfig}. As expected, the sensitivity is
well-controlled by $\Mh$, reflecting the SM-like nature of $h$ for most
points. Indeed, the qualitative behavior discussed above can be
%\marginpar{{\small text added in this and following par}}
understood from the projected search reach for a light SM Higgs in Fig.~\ref{SMfig}. For 
$10 \ifb$ with 10\% analysis improvements the expected limit in the
searches for a light SM-like Higgs at the Tevatron stays below the LEP 
95\% C.L.\ observed limit of $114.4 \gev$~\cite{LEPHiggsSM}. As a consequence,
in the upper left plots of Fig.~\ref{CMSSMfig} and
Fig.~\ref{CMSSMfigmhma} the points outside of the
region of very small $\MA$ and large $\tb$ are either excluded by the
LEP Higgs searches or they show a sensitivity below the $2\sigma$ level.
For 16$\ifb$ and 10\% analysis improvements, the projected SM limit is 
about $119\gev$, while for 10$\ifb$ and 50\% analysis improvements the 
SM limit raises to about $121\gev$. Comparison of the corresponding
plots in Fig.~\ref{CMSSMfig} and Fig.~\ref{CMSSMfigmhma} shows that 
the parameter regions displaying a $2\sigma$ sensitivity are
characterized by $\Mh$ values bounded from above by approximately the projected SM limits. (The MSSM reach is slightly stronger due to the combination with nonstandard channels and the mild increase in the $h\rightarrow b\bar{b}$ branching fraction in the MSSM relative to the SM. These differences also explain why the minimal improvement to achieve 2$\sigma$ coverage in the CMSSM is 30\%, whereas this improvement yields slightly less than 2$\sigma$ significance for a relevant mass range around 125 GeV in the SM.) 
Finally, for 16$\ifb$ and 50\% analysis improvements the search for the
SM Higgs gives rise to a sensitivity above the $2\sigma$ level over the
whole range of $\Mh$ values allowed in the supersymmetric models. This
feature is visible in the lower right plots of 
Fig.~\ref{CMSSMfig} and Fig.~\ref{CMSSMfigmhma}. Furthermore, we find a 
3$\sigma$ median sensitivity for $\Mh\lesssim 116\gev$ in this case.

\begin{figure*}
\begin{minipage}{1.0\linewidth}
\begin{center}
\begin{tabular}{cc}
\includegraphics[width=0.50\textwidth]{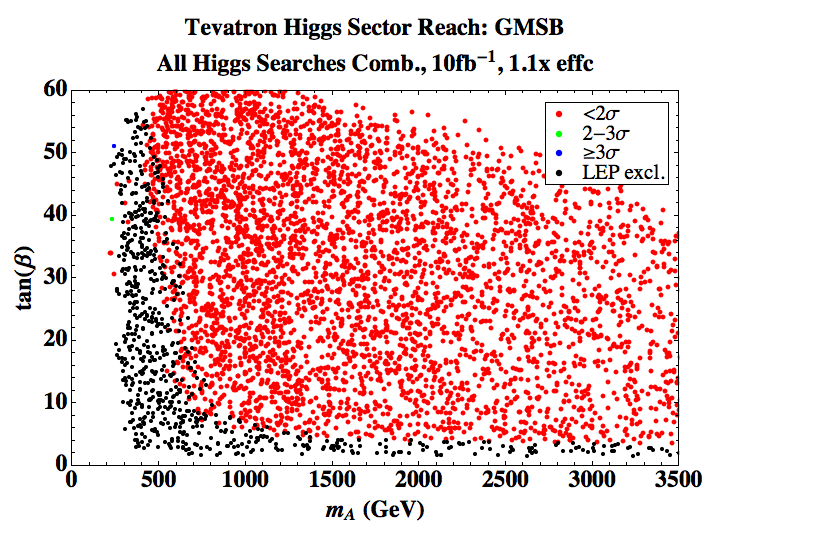} &
\includegraphics[width=0.50\textwidth]{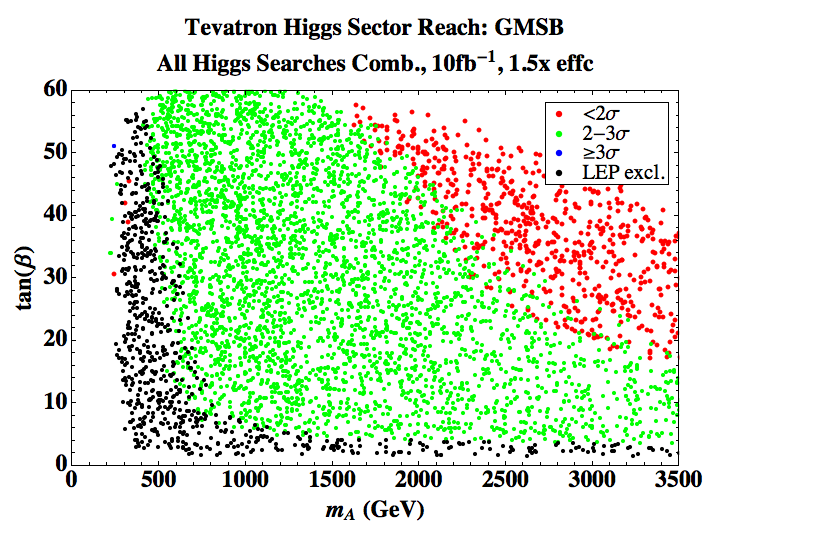}\\
\includegraphics[width=0.50\textwidth]{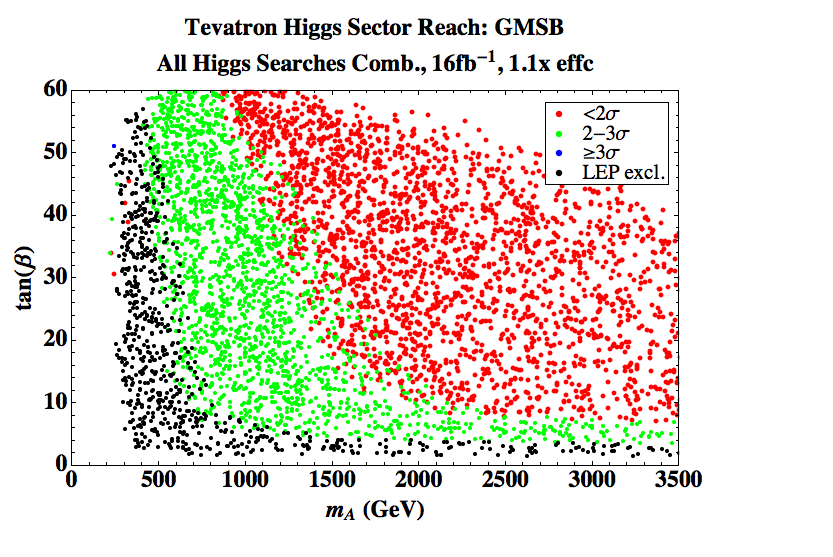} &
\includegraphics[width=0.50\textwidth]{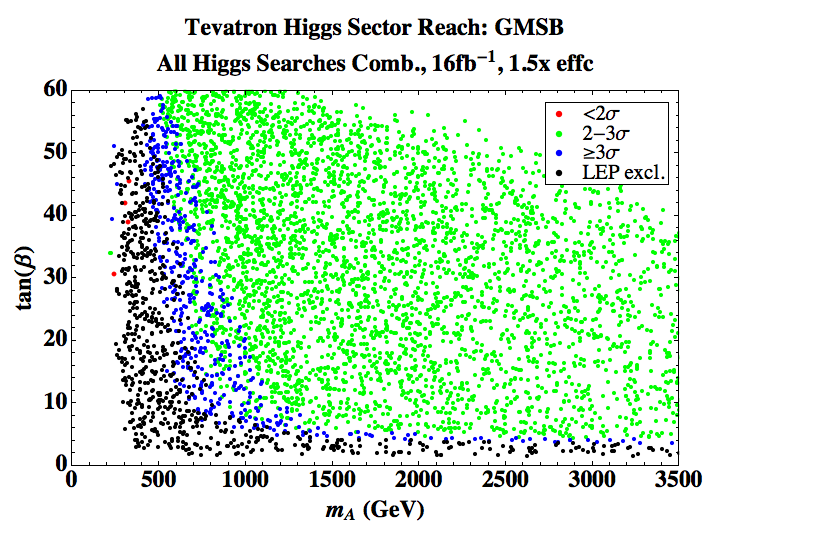}
\end{tabular}
\caption{Projected Tevatron coverage of mGMSB
  on the ($\MA$, $\tb$) plane with $10 \ifb$ and 10\% efficiency
  improvement (\textit{top left}),  $10 \ifb$ and 50\% improvement
  (\textit{top right}), $16 \ifb$ and 10\% efficiency improvement
  (\textit{bottom left}), and $16 \ifb$ with 50\% improvement
  (\textit{bottom right}). The efficiency improvements are given
 relative to the March 2009 expected limits.}\label{GMSBfig} 
\end{center}
\end{minipage}
\end{figure*}

We note that the results of Fig.~\ref{CMSSMfigmhma} allow one
to read off the upper bound on the lightest $\cp$-even Higgs boson mass 
in the CMSSM 
obtained for $m_t = 173.1 \gev$ with state-of-the-art theoretical
predictions. We find an upper bound of $\Mh \leq 125.2 \gev$ in the
CMSSM in the case where $m_t$ is kept fixed and no uncertainties from
unknown higher-order corrections are taken into account.  If the theory and parametric uncertainties are included as
described in Sect.~II, we find an upper value for $m_h$ of $127.4 \gev$. Saturating the bound on $m_h$
would still require an improvement of 30\% in the signal efficiency
to be fully tested at the $2\sigma$ level; the fact that the minimal improvement is the same for both mass values reflects the flatness of the SM projection curves starting near $125\gev$ in Fig.~\ref{SMfig}. Including the theoretical errors could push some points below the $2\sigma$ threshold in the cases of $16\ifb$ with 10\% improvements or $10\ifb$ with 50\% improvements, or below the $3\sigma$ thresholds in the case of $16\ifb$ with 50\% improvements, as is most evident from the results of Fig. 4.  However, the curves in Fig. 1 imply that the theory errors of $\sim 2$~GeV in $m_h$ shift the expected sensitivities by very small amounts ($\sim 0.2\sigma$), and therefore the reach remains close to the sensitivity shown in our figures.
%%%%%%%%%%%%%%%%%%%%%%%%%%%%%%%%%%%%%%%%%%%%%%%%%%%%%%%%%%%%%%%%%%%%%%%%%%%%%%
%%%%%%%%%%%%%%%%%%%%%%%%%%%%%%%%%%%%%%%%%%%%%%%%%%%%%%%%%%%%%%%%%%%%%%%%%%%%%%

%\FloatBarrier

\section{Minimal Gauge-Mediated SUSY-Breaking}

\begin{figure*}
\begin{minipage}{1.0\linewidth}
\begin{center}
\includegraphics[width=0.50\textwidth]{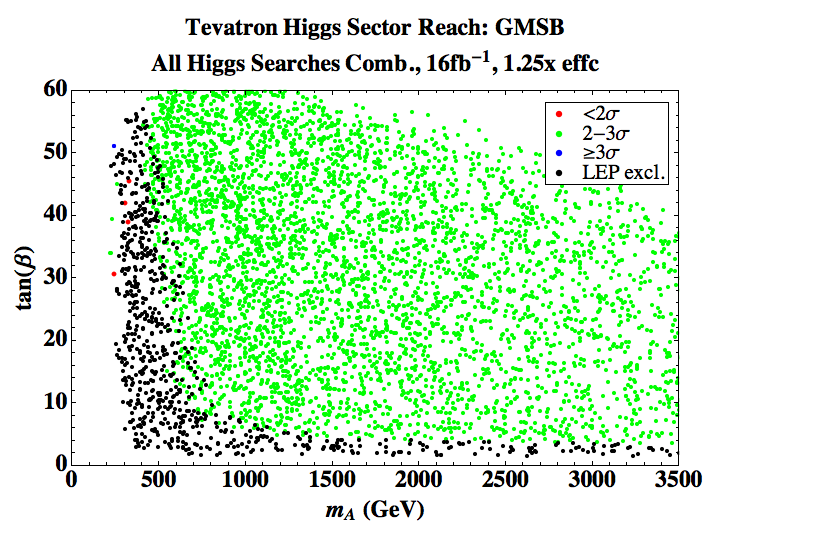}
\caption{Projected Tevatron coverage of mGMSB on the ($\MA$, $\tb$)
  plane with $16 \ifb$ and 25\% efficiency improvement (w.r.t. March
 2009), which is the
  approximate threshold for full coverage at the 2$\sigma$ level. }  
\label{GMSBfig2sig}
\end{center}
\end{minipage}
\end{figure*}

\begin{figure*}
\begin{minipage}{1.0\linewidth}
\begin{center}
\begin{tabular}{cc}
\includegraphics[width=0.50\textwidth]{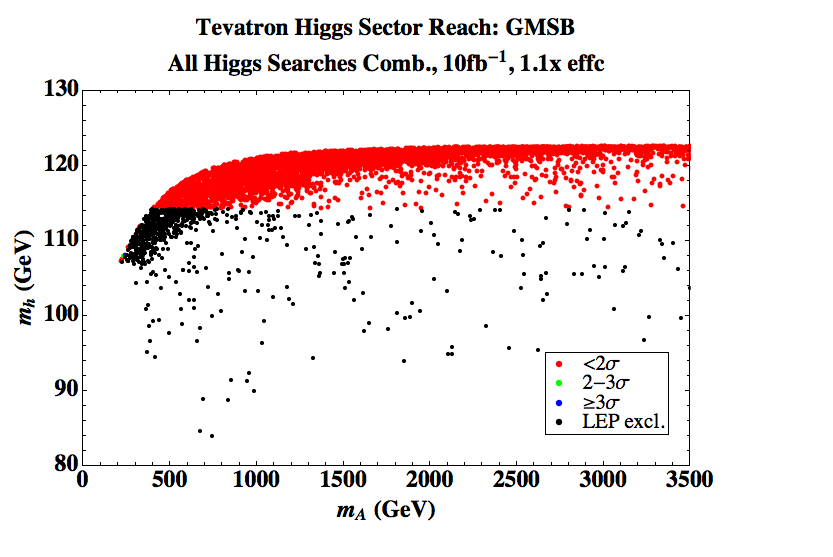} &
\includegraphics[width=0.50\textwidth]{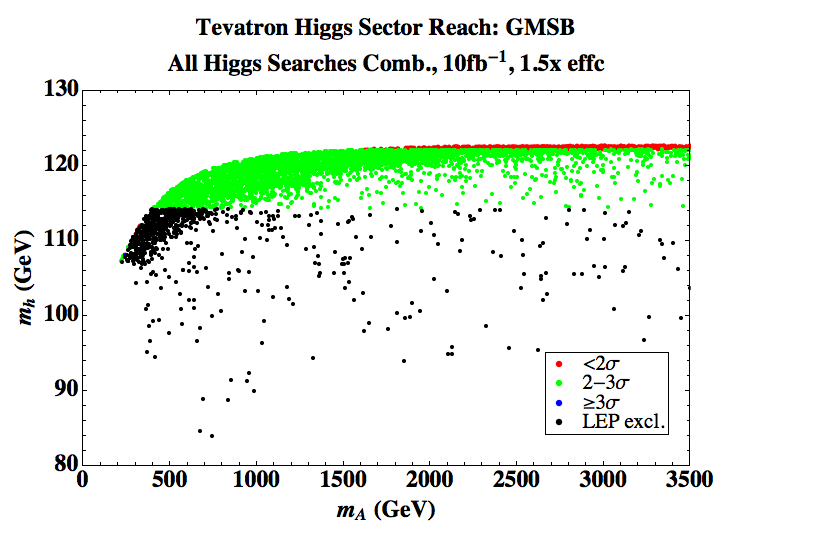} \\
\includegraphics[width=0.50\textwidth]{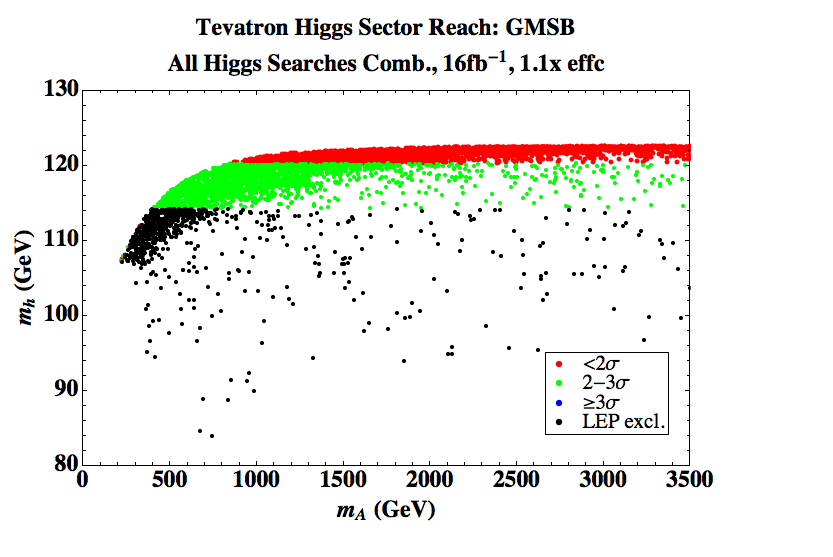} &
\includegraphics[width=0.50\textwidth]{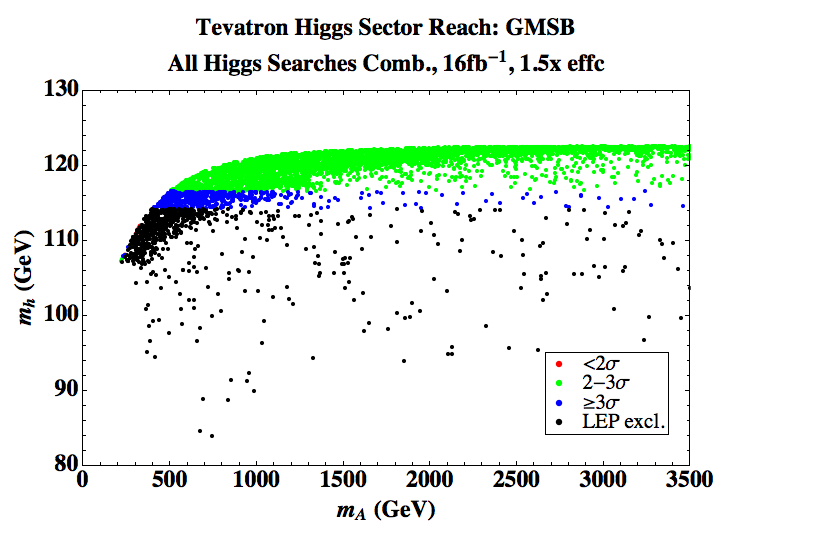}
\end{tabular}
\caption{Projected Tevatron coverage of mGMSB  on the ($\MA$, $\Mh$)
  plane with $10 \ifb$ and 10\% efficiency 
  improvement (\textit{top left}),  $10 \ifb$ and 50\% improvement
  (\textit{top right}), $16 \ifb$ and 10\% efficiency improvement
  (\textit{bottom left}), and $16 \ifb$ with 50\% improvement
  (\textit{bottom right}). The efficiency improvements are given
 relative to the March 2009 expected limits.}\label{GMSBfigmhmA} 
\end{center}
\end{minipage}
\end{figure*}

In the so-called Minimal Gauge Mediation model, the SUSY-breaking
effects are transmitted to the MSSM through 
loops of heavy messenger particles that are charged under the MSSM gauge
groups. The MSSM soft masses are generated by integrating out the
messengers at their mass scale $M_{\rm mess}$, which is not tied to the GUT
scale in general, and evolving the parameters down to the weak scale. At
$M_{\rm mess}$ the soft masses are controlled by gauge couplings, the number
of complete $SU(5)$ $5+\bar{5}$ messenger representations $N_{\rm mess}$,
and a parameter $\Lambda$ which is proportional to the SUSY-breaking
F-term expectation value in the hidden sector.  Soft trilinear couplings
are generated only at higher order and therefore achieve nonzero
values at the weak scale only through RG evolution. We scan the input
parameters in the ranges 
\begin{align}
10 \tev \leq \Lambda \leq 200 \tev, 
&\quad \Lambda \leq M_{\rm mess} \leq 10^5 \times \Lambda, \nonumber\\ 
1 \leq N_{\rm mess} \leq 8, &\quad 1.5 \leq \tb \leq 60.
\end{align}
The Tevatron Higgs reach for the mGMSB scenario is given in Figs.~\ref{GMSBfig}
and~\ref{GMSBfig2sig} on the ($\MA,\tb$) plane.  The panels of
Fig.~\ref{GMSBfig} assume the same increases in luminosity and signal
efficiency as in Fig.~\ref{CMSSMfig}, but Fig.~\ref{GMSBfig2sig} assumes
16~fb$^{-1}$ and a 25\% gain in efficiency. In Fig.~\ref{GMSBfigmhmA} we
project the reach onto the ($\MA,\Mh$) plane with the same parameters
for each panel as in Fig.~\ref{GMSBfig}.
%As in the CMSSM,
%the most important feature is contained in the lower right plot, where  
%$16 \ifb$ and a 50\% increase in efficiency are sufficient to begin
%showing evidence for the SM-like Higgs in some
%cases. Fig.~\ref{GMSBfig2sig}demonstrates that only 25\% improvement is
%necessary to exclude the model at the 95\% C.L. 
%Only here the complete model could be excluded at the 95\% C.L., or it
%would yield at least a $2\sigma$ ``excess'' in the Higgs boson searches.
%Furthermore, 

As explained above, for $10 \ifb$ and no further analysis improvements 
%\marginpar{{\small added text}}
beyond what has already been achieved (i.e., a 10\% improvement compared
to March 2009) the sensitivity of the Tevatron searches for a light
SM-like Higgs boson does not exceed the sensitivity of the LEP Higgs
searches, giving rise to the results displayed in the upper left plots
of Fig.~\ref{GMSBfig} and Fig.~\ref{GMSBfigmhmA} (and in the corresponding
plots for the mAMSB scenario shown below).

For the analyses with $16 \ifb$ and/or further
analysis improvements, two main features emerge. Firstly,
for the upper bound on the lightest $\cp$-even Higgs boson mass in the
mGMSB scenario we find $\Mh \leq 122.6 \gev$ (for $m_t = 173.1 \gev$; the value moves up to $m_h = 124.9 \gev$ if
all theory uncertainties as described in Sect.~II are taken into account),
which is about $2.5\gev$ lower than in the CMSSM, implying a
larger coverage from the Tevatron searches compared to the CMSSM case.
The reduction of the upper bound on $\Mh$
relative to the CMSSM case can be traced
to the stop trilinear coupling, which 
maximizes the radiative contributions to $\Mh$ for 
$A^2_t\approx 4m^2_{\tilde{t}}$
at large $\tb$ (note that the factor of $4$ is due to the on-shell
renormalization scheme used in FeynHiggs; in the MS-bar scheme, the
relation is instead $A^2_t\approx
6m^2_{\tilde{t}}$~\cite{Carena:2000dp}). Because $A_t$ is generated by
2-loop diagrams at the messenger scale in mGMSB, it is typically smaller 
than $m_{\tilde{t}}$ at the electroweak scale. Therefore $\Mh$ tends to
be less than about $120 \gev$, close to the LEP
limit~\cite{mhiggsRG1a,mhiggsRG1,HHH,Ambrosanio:2001xb}.  
The Tevatron reach in the
$b\bar{b}$ channel is thus significant: most points are reached at the
$2\sigma$ level with $10 \ifb$ and 50\% analysis improvements (and all
points are covered at 90\% C.L.).  It should be noted that for
16 fb$^{-1}$, 50\% improvements, 
and $\MA\lesssim 500 \gev$, $3\sigma$ evidence is expected. This can be understood as follows. The direct
dependence of $\Mh$ on $\MA$ is minimal for such large values of $\MA$,
but in mGMSB $\MA$ and the squark masses are both controlled by $\Lambda$. Consequently,
%\marginpar{{\small GW: I find this unclear; what is really meant here? PD: rephrased.}}
lower values of $\MA$ are typically correlated with lighter squarks.  In addition to the reduction in $\Mh$ from small $A_t$, the lower stop masses in this region of parameters also reduce
$\Mh$, so that the Tevatron Higgs searches have a high sensitivity.
%for covering the mGMSB parameter space.
Again we note
that as with the CMSSM, it is only with a combination of increased
luminosity and refinements in the signal extraction that $3\sigma$
sensitivity is obtained. However, we stress that, as shown in
%\marginpar{{\small GW: rephrased}}
Fig.~\ref{GMSBfig2sig}, only 15\% improvement is necessary 
(i.e., 25\% improvement w.r.t.\ March 2009) to achieve the sensitivity to
exclude any model point at the 95\% C.L.\ with $16 \ifb$
and therefore to completely
rule out this widely studied SUSY scenario. Increasing $m_h$ by the theory
uncertainties discussed in Section II requires an improvement of 20\% (30\% w.r.t.\ March 2009) in the
signal efficiency to fully cover mGMSB at $2\sigma$.

\begin{figure*}
\begin{minipage}{1.0\linewidth}
\begin{center}
\begin{tabular}{cc}
\includegraphics[width=0.50\textwidth]{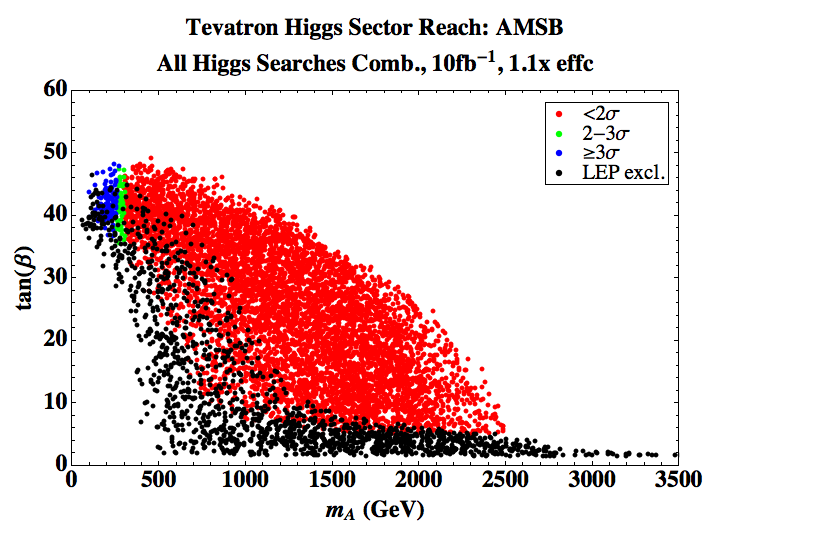} &
\includegraphics[width=0.50\textwidth]{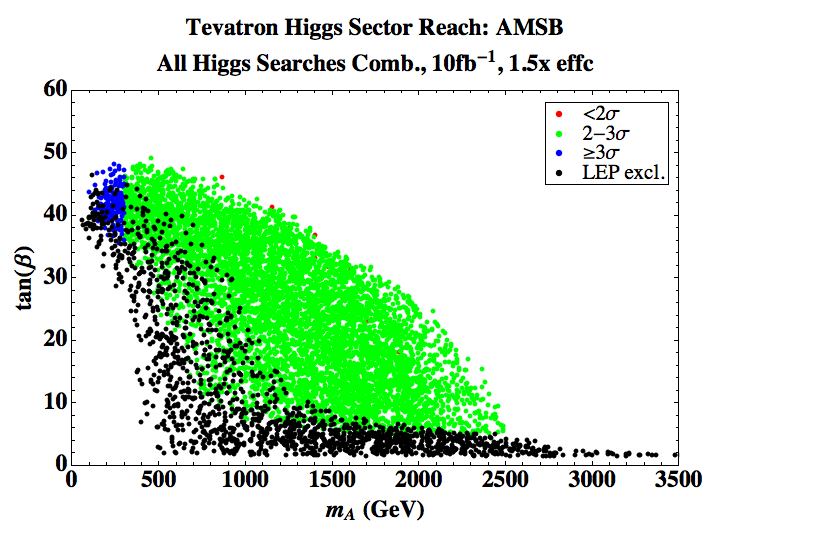}\\
\includegraphics[width=0.50\textwidth]{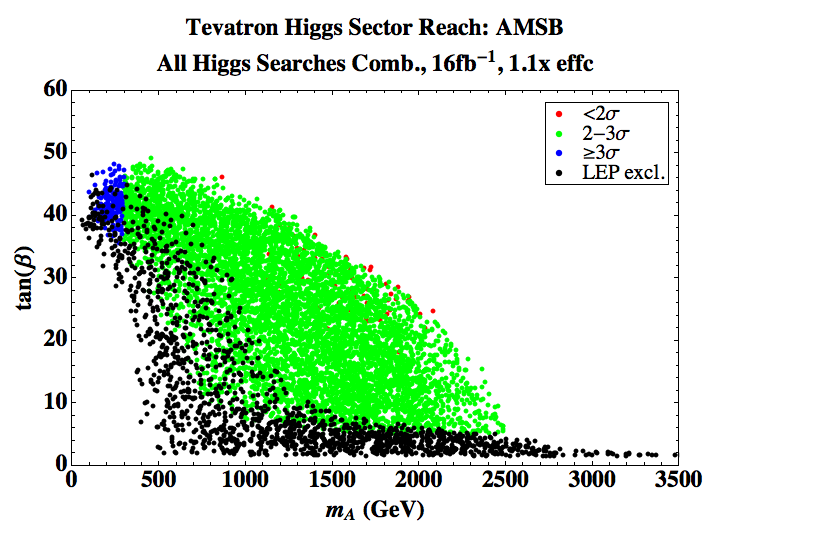} &
\includegraphics[width=0.50\textwidth]{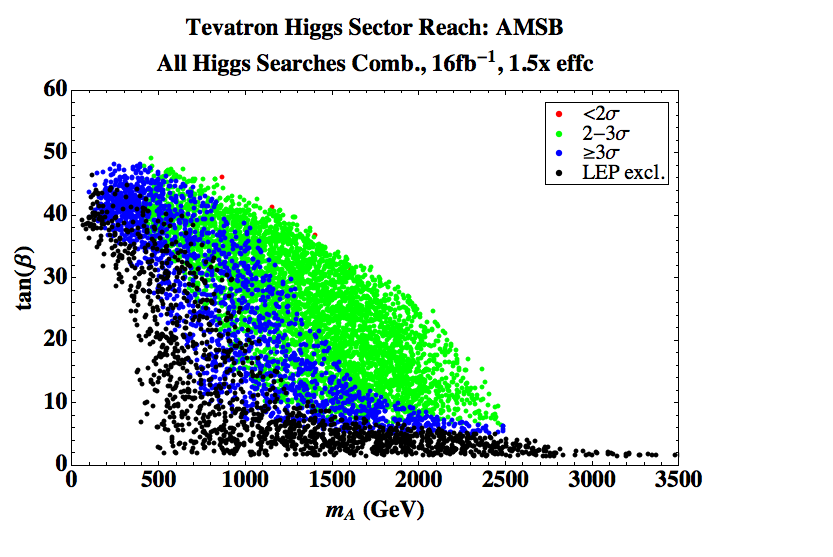}
\end{tabular}
\caption{Projected Tevatron coverage of mAMSB on the
  ($\MA$, $\tb$) plane with $10 \ifb$ and 10\% efficiency improvement
  (\textit{top left}),  $10 \ifb$ and 50\% improvement (\textit{top
    right}), $16 \ifb$ and 10\% efficiency improvement (\textit{bottom
    left}), and $16 \ifb$ with 50\% improvement (\textit{bottom
    right}). The efficiency improvements are given
 relative to the March 2009 expected limits.}
    \label{AMSBfig} 
\end{center}
\end{minipage}
\end{figure*}

The second important feature apparent in our analysis for the
mGMSB model is the absence of a $3\sigma$ reach in the
nonstandard channels at large $\tb$ and low $\MA$. More precisely, such
points do not arise at all in the scan, because they are associated with
slepton masses below the experimental limit.  Unlike in the CMSSM, the
boundary values for 
the slepton masses are suppressed relative to the squarks by factors of
the electroweak gauge couplings, and at large $\tb$ and low $\Lambda$,
the $\tau$ Yukawa coupling 
is sufficiently enhanced to drive the squared slepton
masses to small or even negative values at the electroweak scale.  Thus
it can be expected that 
%\marginpar{{\small GW: I don't agree with this statement; remove? PD: changed supersymmetric to mGMSB}}
the primary signal of an mGMSB Higgs sector at the Tevatron (with $16 \ifb$ and $50\%$
 analysis improvements) will
be $2-3\sigma$ evidence for an SM-like state in the associated production
channel with $m_h<125 \gev$ (including in this upper bound possible contributions from both the theoretical and parametric uncertainties.)

%\begin{figure*}[t]
%\begin{minipage}{1.0\linewidth}
%\hspace{-1.5cm}
%\begin{center}
%\begin{tabular}{cc}
%\includegraphics[width=0.50\textwidth]{GMSB_mhtb_lin_10ifb_1pt1x} &
%\includegraphics[width=0.50\textwidth]{GMSB_mhtb_lin_10ifb_1pt5x} \\
%\includegraphics[width=0.50\textwidth]{GMSB_mhtb_lin_16ifb_1pt1x} &
%\includegraphics[width=0.50\textwidth]{GMSB_mhtb_lin_16ifb_1pt5x}
%\end{tabular}
%\caption{Projected Tevatron coverage of minimal gauge mediation models
%  on the ($\tb$, $\Mh$) plane with $10 \ifb$ and 10\% efficiency
%  improvement (\textit{top left}),  $10 \ifb$ and 50\% improvement
%  (\textit{top right}), $16 \ifb$ and 10\% efficiency improvement
%  (\textit{bottom left}), and $16 \ifb$ with 50\% improvement
%  (\textit{bottom right}).}\label{GMSBfigmhtb} 
%\end{center}
%\end{minipage}
%\end{figure*}

%%%%%%%%%%%%%%%%%%%%%%%%%%%%%%%%%%%%%%%%%%%%%%%%%%%%%%%%%%%%%%%%%%%%%%%%%%%%%%
%%%%%%%%%%%%%%%%%%%%%%%%%%%%%%%%%%%%%%%%%%%%%%%%%%%%%%%%%%%%%%%%%%%%%%%%%%%%%%

%\FloatBarrier
%\clearpage
\section{Anomaly-Mediated SUSY-breaking}

SUSY-breaking in the simplest phenomenologically acceptable realization
of the mAMSB scenario 
is governed by two parameters: an F-term $M_{\rm aux}$, to which all
MSSM soft parameters are proportional, and an explicit universal
sfermion mass $m_0$. The soft masses at all scales are given by simple
functions of the gauge and Yukawa $\beta-$functions and the anomalous
dimensions of the fields. We scan in the ranges 
\begin{align}
0 \leq m_0 \leq 2 \tev, &\quad 20 \tev \leq M_{\rm aux}\leq 100 \tev,\nonumber\\
&\quad 1.5 \leq \tb \leq 60.
\end{align}

\begin{figure*}
\begin{minipage}{1.0\linewidth}
\begin{center}
\includegraphics[width=0.50\textwidth]{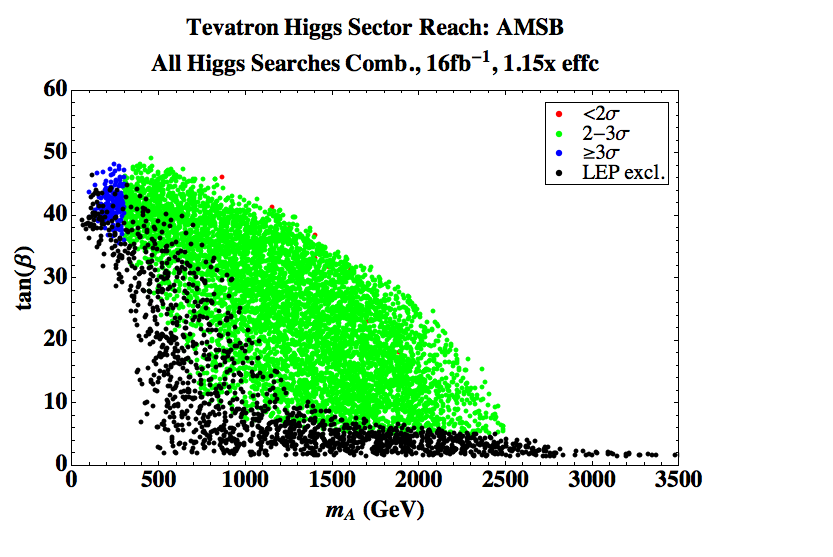}
\caption{Projected Tevatron coverage of mAMSB on the ($\MA$, $\tb$)
  plane with $16 \ifb$ and 15\% efficiency improvement (w.r.t. March
 2009), which is the
  approximate threshold for full coverage at the 2$\sigma$ level.}  
\label{AMSBfig2sig}
\end{center}
\end{minipage}
\end{figure*}

\begin{figure*}
\begin{minipage}{1.0\linewidth}
\begin{center}
\begin{tabular}{cc}
\includegraphics[width=0.50\textwidth]{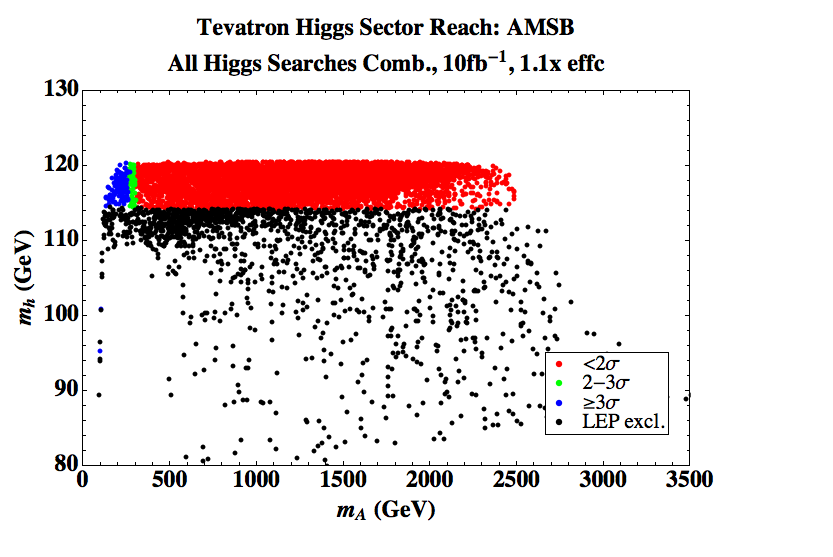} &
\includegraphics[width=0.50\textwidth]{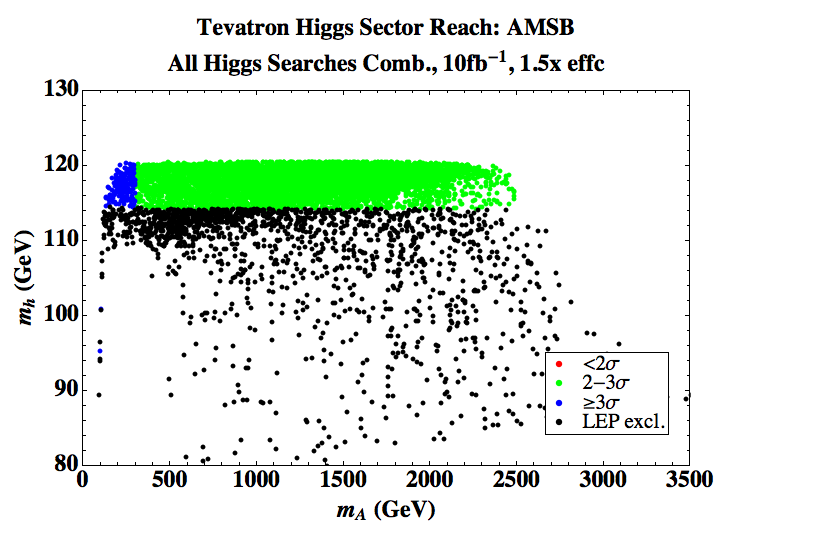} \\
\includegraphics[width=0.50\textwidth]{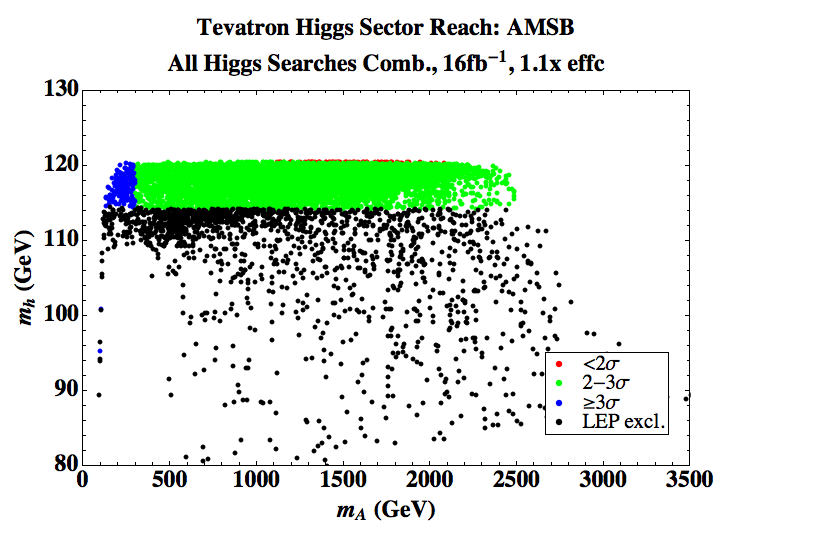} &
\includegraphics[width=0.50\textwidth]{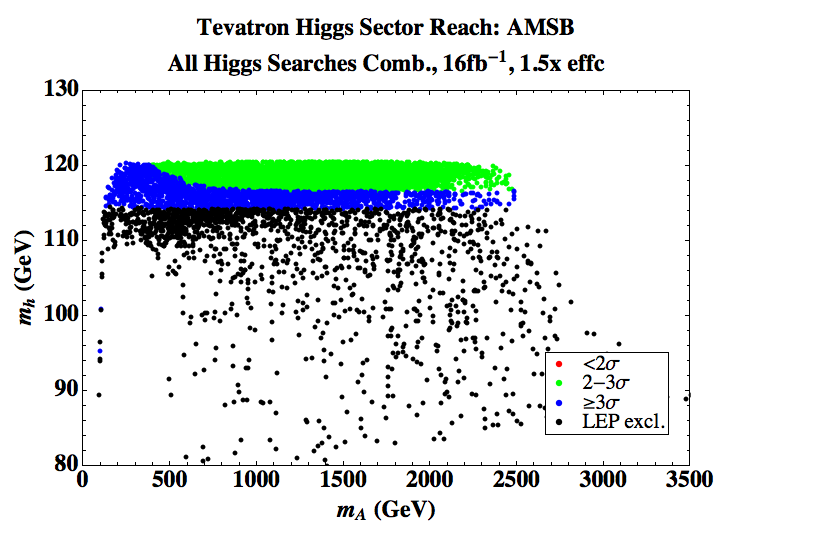}
\end{tabular}
\caption{Projected Tevatron coverage of mAMSB on the
  ($\MA$, $\Mh$) plane with $10 \ifb$ and 10\% efficiency improvement
  (\textit{top left}),  $10 \ifb$ and 50\% improvement (\textit{top
    right}), $16 \ifb$ and 10\% efficiency improvement (\textit{bottom
    left}), and $16 \ifb$ with 50\% improvement (\textit{bottom
    right}). The efficiency improvements are given
 relative to the March 2009 expected limits.}\label{AMSBfigmhma} 
\end{center}
\end{minipage}
\end{figure*}

The Higgs sector reach is given in Figs.~\ref{AMSBfig}
and~\ref{AMSBfig2sig} on the ($\MA,\tb$) plane. As before, the panels of
Fig.~\ref{AMSBfig} assume increases in luminosity and signal efficiency
as in Fig.~\ref{CMSSMfig}, but Fig.~\ref{AMSBfig2sig} assumes 16
fb$^{-1}$ and a 15\% gain in efficiency (i.e., about a 5\% gain w.r.t.
the present situation). 
%\marginpar{{\small GW: added comment}}
The reach on the ($\MA,\Mh$)
plane is given in Fig.~\ref{AMSBfigmhma} with the same parameters for
each panel as in Fig.~\ref{AMSBfig}. 
%\marginpar{{\small GW: added text}}
For the upper bound on the lightest $\cp$-even Higgs boson mass in the
mAMSB scenario we find $\Mh \leq 120.5 \gev$ (for $m_t = 173.1 \gev$; the value moves up to $\Mh = 122.5 \gev$ if
all theory uncertainties as described in Sect.~II are taken into account), 
which is about $2.1\gev$ lower than in the mGMSB case, and 
about $4.7\gev$ lower than in the CMSSM case. Thus, among the scenarios
considered here, the potential of the Tevatron Higgs searches to
completely cover the whole parameter space of the model will be highest
in the mAMSB scenario. Accordingly, for $16 \ifb$ and as little as 5\%
increase in efficiency (i.e., 15\% increase as compared to March 2009)
there will be 2$\sigma$ exclusion power for all scan points.  Taking into account the theory
uncertainties for $\Mh$ discussed above requires an increase of 25\% w.r.t. 2009
in the experimental efficiency to fully cover mAMSB at 2$\sigma$.
For $16 \ifb$ and a 50\% increase in efficiency w.r.t. 2009,
there is a broad potential for 3$\sigma$ evidence.
The former is demonstrated in
Fig.~\ref{AMSBfig2sig} (with the exception of points at the boundary of
the parameter space that will be discussed below).  
%Only here
%the complete model
%could be excluded at the 95\% C.L., or it would yield at least a
%$2\sigma$ ``excess'' in the Higgs boson searches (with the exception of
%very few points that will be discussed below).

The Tevatron reach can be
understood from arguments similar to the previous two models.  In the
absence of $m_0$ the sleptons would always be tachyonic, but the
introduction of this parameter allows positive masses squared even for
large $\tb$. Therefore, as in the CMSSM, the sensitivity to the
nonstandard Higgs in the $\tau\tau$ and 3$b$ channels is high for low
$\MA$.  
The reduced upper bound on $\Mh$ as compared to the CMSSM and the mGMSB
scenario
is attributable to the stop trilinear
coupling, which is proportional to $M_{\rm aux}$ and the
$\beta$-function of the top Yukawa coupling, 
and is generically of the same order
or smaller than $m_{\tilde{t}}$.  As a consequence of the relatively low 
upper bound on $\Mh$, a large fraction of the parameter space is
reachable at 2$\sigma$ level
with luminosity gains alone, and the $3\sigma$ reach for the light Higgs
is significant with $16 \ifb$ and 50\% improvements.  

However, near the
largest values of $m_0$ and smallest values of $M_{\rm aux}$, a few points 
avoid even $2\sigma$ sensitivity. In these models the squarks
are heavy, pushing up $\Mh$, while the gaugino masses are light, opening
the decay channel $h\to\tilde{\chi}^0_1\tilde{\chi}^0_1$. Although we
did not treat this case specially in our projections, the analysis of
Ref.~\cite{Davoudiasl:2004aj} indicates that evidence for an
invisibly-decaying SM-like Higgs may be achievable at the Tevatron by
combining searches in the weak boson fusion and associated production
channels with $12 \ifb$, 
and that discovery may occur at the LHC with $10\ifb$ and 
$\sqrt{s} = 14 \tev$ in the $Zh$ channel alone.

%\begin{figure*}[t]
%\begin{minipage}{1.0\linewidth}
%\hspace{-1.5cm}
%\begin{center}
%\begin{tabular}{cc}
%\includegraphics[width=0.50\textwidth]{AMSB_mhtb_lin_10ifb_1pt1x} &
%\includegraphics[width=0.50\textwidth]{AMSB_mhtb_lin_10ifb_1pt5x} \\
%\includegraphics[width=0.50\textwidth]{AMSB_mhtb_lin_16ifb_1pt1x} &
%\includegraphics[width=0.50\textwidth]{AMSB_mhtb_lin_16ifb_1pt5x}
%\end{tabular}
%\caption{Projected Tevatron coverage of anomaly mediation models on the
%  ($\tb$, $\Mh$) plane with $10 \ifb$ and 10\% efficiency improvement
%  (\textit{top left}),  $10 \ifb$ and 50\% improvement (\textit{top
%    right}), $16 \ifb$ and 10\% efficiency improvement (\textit{bottom
%    left}), and $16 \ifb$ with 50\% improvement (\textit{bottom
%    right}).}\label{AMSBfigmhtb} 
%\end{center}
%\end{minipage}
%\end{figure*}

%%%%%%%%%%%%%%%%%%%%%%%%%%%%%%%%%%%%%%%%%%%%%%%%%%%%%%%%%%%%%%%%%%%%%%%%%%%%%%
%%%%%%%%%%%%%%%%%%%%%%%%%%%%%%%%%%%%%%%%%%%%%%%%%%%%%%%%%%%%%%%%%%%%%%%%%%%%%%

%\FloatBarrier
\section{Conclusions}

%In this note we have attempted to quantify the physics potential of the
%Tevatron collider in the context of the MSSM Higgs sector, based on 
%\marginpar{{GW: extended, rephrased}}
%$10 \ifb$ or $16 \ifb$ of analyzable data per experiment, corresponding
%to Tevatron operation until the end of 2011 or 2014, respectively.
%We have investigated the most commonly considered 
%high-scale models for the communication of SUSY-breaking to the MSSM, 
%in which predictions for the low-scale spectra and the collider reach
%are given in terms of relatively few free parameters. 
%As a result, we have provided
%projections for the Constrained MSSM, the minimal gauge mediation of
%SUSY-breaking, and anomaly mediation. 

In this note we have analyzed the physics potential of the Tevatron collider in the context of the MSSM Higgs sector, based on $10 \ifb$ or $16 \ifb$ of analyzable data per experiment, corresponding to Tevatron operation until the end of 2011 or 2014, respectively. For the projections we have also studied the impact of possible improvements in the efficiencies of the Tevatron analyses (an extended
 running time and higher accumulated statistics should of course be
 helpful for achieving such efficiency improvements, as some uncertainties in Higgs analyses can be sensitive to statistical uncertainties in other measurements).
%In this article we have analyzed the physics potential of an extended Tevatron run, beyond the
%projected shutdown at the end of the year 2011.  An extension of the Tevatron run will not
%only provide an increase in statistics but also will help to realize the efficiency improvements
%needed to obtain evidence of a Higgs boson in the relevant mass region below 140~GeV.
%We have worked in the context of the MSSM, based on 10 fb$^{-1}$ and 16 fb$^{-1}$ corresponding
%to Tevatron operation until the end of 2011 or  2014, and analyzed the Tevatron Higgs reach in different efficiency improvement
%scenarios. 
We have investigated the most commonly considered 
high-scale models for the communication of SUSY-breaking to the MSSM, 
in which predictions for the low-scale spectra and the collider reach
are given in terms of relatively few free parameters. 
As a result, we have provided
projections for the Constrained MSSM, the minimal gauge mediation of
SUSY-breaking, and minimal anomaly mediation. 

For $10 \ifb$, i.e.\ Tevatron running until the end of 2011, and no 
further improvements in the analysis efficiency compared to the present
situation, the sensitivity of the Tevatron searches for a light SM-like
Higgs would not exceed the sensitivity of the Higgs searches at LEP. 
Thus, the impact of the Tevatron Higgs searches in the different SUSY
scenarios would be rather limited in this case, with the best prospects 
in the parameter region of small $\MA$ and very large $\tb$
for nonstandard Higgs searches in the
$\tau\tau$ and $3b$ channels. 

If $16 \ifb$ can be analyzed at each Tevatron experiment, as expected 
from running the Tevatron for three additional years beyond 2011, and the analysis
efficiency can be 
improved by 30\% with respect to the status of March 2009 (where 10\%
between March 2009 and summer 2010 has been realized already), a
$2\sigma$ (or higher) sensitivity is 
expected over the whole parameter space in all three
models. Consequently, all three different types of SUSY-breaking models
considered here 
could be excluded at the 95\% C.L., or would yield at least a
$2\sigma$ ``excess'' in the Higgs boson searches.
It should be noted that an exclusion of those SUSY scenarios, which up to
now have been used for defining the benchmarks for SUSY searches at the 
LHC and elsewhere, could have profound consequences on the possible
interpretation of SUSY searches at the LHC.

With an integrated luminosity of $16 \ifb$ per experiment and 
a 50\% improvement in the signal efficiency with respect to the status
of March 2009, the opportunity for $3\sigma$
evidence for a SM-like Higgs will open up in significant parts of the
parameter space of the most prominent SUSY-breaking scenarios.
%Similarly $2\sigma$ exclusions are expected for essentially all models, barring
%unconventional Higgs decay channels.  
We also stress that our projections could be considered conservative in the sense that we have used a flat efficiency improvement profile that is weaker in the low-mass region than the improvements projected by the Tevatron collaborations. Using the efficiency improvements presented in~\cite{denisovPAC}
would result in widespread $3\sigma$ coverage in all of the models
we considered for $16 \ifb$ of data.
%Using the stronger improvement scenario would result in widespread $3\sigma$ coverage in all of the models we considered for $16 \ifb$ of data.

Perhaps the most exciting possibility is that 
%the Tevatron and LHC would simultaneously
%see signals in their respective search channels. 
indications of a Higgs signal would build up simultaneously at the Tevatron and LHC in their respective search channels.
The $gg\rightarrow h \rightarrow \gamma \gamma$ channel
will eventually provide a discovery channel at the LHC  and will allow a precise mass measurement, 
while the simultaneous observation of the Higgs in the associated production channel at the
Tevatron will strengthen the link to EWSB and will provide direct information on its coupling
to weak gauge bosons and to bottom quarks.  The combination of the LHC and Tevatron channels can thus be important input for a Higgs boson coupling determination. The combined significance of the LHC and Tevatron channels
can also be helpful in the observation of a Higgs in the
near future in cases in which the production cross sections in one or
more of the channels are suppressed with respect to the SM expectations.\\
\\

%The combined
%significance of both channels can also be helpful in the observation of a Higgs in the
%near future in cases in  which the production cross sections in one or both relevant channels 
%are suppressed with respect to the SM expectations.

%{\bf GW: add comment that increased running time will enhance prospects
%for analysis improvements?}

%%%%%%%%%%%%%%%%%%%%%%%%%%%%%%%%%%%%%%%%%%%%%%%%%%%%%%%%%%%%%%%%%%%%%%%%%%%%%%
%%%%%%%%%%%%%%%%%%%%%%%%%%%%%%%%%%%%%%%%%%%%%%%%%%%%%%%%%%%%%%%%%%%%%%%%%%%%%%

\begin{acknowledgements}
Work supported in part by the European Community's Marie-Curie Research
Training Network under contract MRTN-CT-2006-035505 `Tools and Precision Calculations for Physics Discoveries at Colliders'.  Fermilab is operated by Fermi Research Alliance, LLC under Contract No. DE-AC02-07CH11359 with the U.S. Department of Energy. Work at ANL is supported in part by the U.S. Department of Energy (DOE), Div.~of HEP, Contract DE-AC02-06CH11357. This work was supported in part by the DOE under Task TeV of contract DE-FGO2-96-ER40956.  T.~L. is supported by the Fermi-McCormick Fellowship and the DOE Grant DE-FG02-91ER40618 at U. California, Santa Barbara.  M.~C. and C.~W. would like to thank the Aspen Center for Physics, where part of this work was completed.
\end{acknowledgements}
%\newpage

\FloatBarrier

\end{document}